\documentclass[12pt]{article}
\pdfoutput=1
\usepackage{amsbsy}
\usepackage{amsfonts}
\usepackage{amsmath,amssymb}
\usepackage{bbold}
\usepackage{graphicx}

\newcommand{\eq}{\begin{equation}}
\newcommand{\eqa}{\begin{eqnarray}}  
\newcommand{\en}{\end{equation}}
\newcommand{\ena}{\end{eqnarray}}
\newcommand{\enn}{\nonumber \end{equation}}

\def\sk{\vskip .4cm}
\def\noi{\noindent}

\def\al{\alpha}
\def\be{\beta}

\let \part\partial

\def\part{\partial}

\def\sk{\vskip .4cm}

\def\noi{\noindent}

\def\X0{X^0}

\def\al{\alpha}

\def\square{{\,\lower0.9pt\vbox{\hrule \hbox{\vrule height 0.2 cm
\hskip 0.2 cm \vrule height 0.2 cm}\hrule}\,}}

\def\lb{\langle}
\def\rb{\rangle}

\begin{document}

\begin{titlepage}

\vskip 2em
\begin{center}
{\Large \bf No relation for Wigner's friend} \\[3em]

\vskip 0.5cm

{\bf
Leonardo Castellani}
\medskip

\vskip 0.5cm

{\sl Dipartimento di Scienze e Innovazione Tecnologica
\\Universit\`a del Piemonte Orientale, viale T. Michel 11, 15121 Alessandria, Italy\\ [.5em] INFN, Sezione di 
Torino, via P. Giuria 1, 10125 Torino, Italy\\ [.5em]
Arnold-Regge Center, via P. Giuria 1, 10125 Torino, Italy
}\\ [4em]
\end{center}

\begin{abstract}
\sk

We argue that Wigner's friend thought experiment does not support observer dependence of quantum states. An analysis in terms of history vectors suggests that quantum collapse is to be understood as collapse of histories rather than collapse of states.

\end{abstract}

\vskip 10cm\
 \noi \hrule \vskip .2cm \noi {\small
leonardo.castellani@uniupo.it}

\end{titlepage}

\newpage
\setcounter{page}{1}



The core of subjective or relational interpretations of QM  (see for ex. \cite{Qbism1,Qbism2,Everett,Rovelli}) is the alleged {\sl observer dependence} of quantum states, meaning that different observers give different descriptions of the same quantum system. This is claimed by some authors to be a consequence of the usual quantum mechanical rules, when these are applied to ``third person" situations, as in Wigner's friend thought experiment \cite{Wigner1961}. In the present note we argue that the usual textbook QM description of the experiment does not lead to an essential observer dependence. 

Let us briefly recall the setting: an observer $F$ and a system $S$ are contained in a perfectly isolated laboratory. A second observer $W$ stays outside of the laboratory. The initial state of $S$ at $t=t_0$ is prepared in the superposition
 \eq
 \al |0\rb + \be |1\rb
 \en
At $t=t_1$ the observer $F$ effects a measurement on $S$ in the $|0\rb,|1\rb$ basis \footnote{i.e. a measurement of an observable having $|0\rb$ and $|1\rb$ as eigenvectors.}.
According to the usual QM rules, $F$ obtains the result 0 with probability $|\al|^2$ and the result 1 with probability $|\be|^2$. Suppose for example that $F$ obtains 1: ``for him" the initial state $|\psi\rb$ collapses into the state $|1\rb$:
\eq
\al |0\rb + \be |1\rb \longrightarrow |1\rb \label{evolutionS}
\en

How does the external observer $W$ describe the evolution, from $t_0$ to $t_1$, of the composite system $S+F$ ?  According to QM, in absence of any interaction with the environment, the system $S+F$ evolves unitarily. The initial state of the composite system
is the tensor product $(\al |0\rb + \be |1\rb) \otimes |init\rb$, where $|init\rb$ is the initial state of the observer $F$. A measurement by $F$ on $S$ is modeled by the $S+F$ (unitary) evolution
\eqa
& & |0\rb  \otimes |init\rb \longrightarrow |0\rb  \otimes |0_F\rb \label{m0}\\
& & |1\rb  \otimes |init\rb \longrightarrow |1\rb  \otimes |1_F\rb \label{m1}
\ena
where $|0_F\rb$  (resp. $|1_F\rb$) is the state of $F$ after he has obtained 0 (resp. 1), or equivalently
the state of a notebook where $F$ writes down the result of the measurement. 
\sk
The crucial point is that
the state of $S$ becomes correlated to the state of $F$. Thus if the initial state of $S+F$ is
$(\al |0\rb + \be |1\rb) \otimes |init\rb$, by linearity the system evolves from $t_0$ to $t_1$ as
\eq
(\al |0\rb + \be |1\rb) \otimes |init\rb \longrightarrow \al |0\rb \otimes |0_F\rb  + \be |1\rb  \otimes |1_F\rb
\label{evolution2}
\en
i.e. in an {\sl entangled} state of the $S+F$ system. All predictions on future measurements by $W$ 
must be computed using this entangled state. In particular $W$ could observe 
interference effects in $S+F$, since the state at $t_1$ is a superposition. 

The observer $F$, on the contrary, "sees" no superposition. In fact
he cannot observe $S+F$, since he is {\sl part} of the composite system. Only $S$ is accessible to him, and
after his measurement at $t=t_1$, $F$ describes the state of $S$ as one of the two states $|0\rb$, $|1\rb$.

So far no difficulty arises. $W$ and $F$ are describing different physical systems: no wonder that
their descriptions differ. 

But suppose that $W$ focuses his attention on the subsystem $S$. He can make measurements on $S$, since $S$ is a subsystem of $S+F$ and therefore accessible to him. How does $W$ describe the state of $S$ after $t_1$, before making any measurement himself? Is this description different from the one made by $F$ ? Here the question acquires relevance, since both observers refer to the same system $S$.
In fact it lies at the heart of the {\sl measurement problem} in QM, highlighting the clash between unitary evolution of isolated systems and nonunitary projection due to measurement.

The answer of textbook QM involves the reduced density operator  $\rho^{(S)}$ for system $S$. The density operator $\rho^{(S+F)}$ for $S+F$ at $t=t_1$ is given by:
\eq
\rho^{(S+F)} =  (\al |0\rb \otimes |0_F\rb  + \be |1\rb  \otimes |1_F\rb)( \al^\star  \lb 0| \otimes \lb 0_F|  + \be^\star \lb 1 |  \otimes \lb 1_F|)
\en
The reduced density operator $\rho^{(S)}$ is defined by tracing on the subsystem $F$. Assuming
$|0_F\rb$ and $|1_F\rb$ to be orthogonal\footnote{since they are eigenvectors of a ``position of the pointer" observable.} we find
\eq
\rho^{(S)} = Tr_F ( \rho^{(S+F)} ) = |\al |^2 |0\rb \lb 0| + |\be |^2 |1\rb \lb 1| 
\en
Thus $W$ describes $S$ to be in a {\sl mixed} state, and precisely in the state $|0\rb$ with
probability $|\al |^2$ and in the state $|1\rb$ with
probability $|\be |^2$. Note that this mixed state is radically different from the superposed (pure) state
$ \al |0\rb + \be |1\rb$. It is the description of system $S$ given by $W$, after a measurement has been performed by $F$, but without acquisition of the result (since $S+F$ is isolated, $F$ cannot communicate with $W$).
\sk
In summary, the $F$ and $W$ descriptions of the state of $S$ at $t_0$ and $t_1$ can be represented as follows:
\eqa
 & & ~~~~~~~~~~~~~~t_0~~~~~~~~~~~~~~~~~~~~~~~~~~~~~~~~~~~~~~~t_1 \nonumber \\
 & & F:~~~\al |0\rb + \be |1\rb  \longrightarrow~ |0\rb ~or~ |1\rb~~with~ probabilities ~|\al|^2~ or~ |\be|^2 \nonumber \\
 & & W:~~~\al |0\rb + \be |1\rb  \longrightarrow~ mixed~ state~ (statistical ~ensemble ~of ~ |0\rb ~and~ |1\rb)   \nonumber \\
 & & ~~~~~~~~~~~~~~~~~~~~~~~~~~~~~~~with~ probabilities ~|\al|^2~ and~ |\be|^2 \nonumber
 \ena
The two descriptions do not coincide, because $F$ knows the result of his measurement, whereas $W$ does not. This is the only reason of the difference, due to ignorance of $W$, and reflected in the statistical ensemble he must use to describe $S$ at $t_1$. But both $F$ and $W$ do agree that the initial state of 
$S$ has collapsed in one of the two basis states $|0\rb$ or $|1\rb$, even if only $F$ knows which one.
This difference has no profound significance: $F$ would describe $S$ in exactly the same way as
$W$, if he chose to measure $S$ (producing collapse) without registering the result. Moreover the difference is not due to quantum mechanical effects, but only to incomplete information of one of the two observers. The same difference arises in a classical world, where $S$ is for example a coin under a cup, and $F$ ``measures" it by removing the cup and registering whether $S$ shows head or tails.
If $W$ cannot communicate with $F$ ($S+F$ being isolated), the information on the result of $F$  measurement is not available, and he can only describe the state of $S$ statistically, with a 50-50 probability of heads or tails. But this could hardly be a motivation to develop a relational interpretation of classical mechanics\footnote{besides the obvious system of reference dependence of position, velocities etc.}.

\sk
To acquire the missing information, $W$ must interact with the $S+F$ system, for example with a measurement on $S$, or on $F$, or on the whole $S+F$. If he interacts with the whole of $S+F$,
he will be able to detect that the composite system is in the superposition state (\ref{evolution2}). But if he limits his measurements to the subsystems $S$ or $F$, these are not in superposed (pure) states, but in mixed states, and no interference effects can be detected. 

Suppose that $W$ measures $S$ and obtains the result 0. The $S+F$ system collapses:
\eq
\al |0\rb \otimes |0_F\rb  + \be |1\rb  \otimes |1_F\rb \longrightarrow |0\rb  \otimes |0_F\rb
\en
and by a subsequent measure on $F$, $W$ can verify that 0 was indeed the result registered by $F$.
Thus no contradiction can arise between the results of $F$ and $W$. 

But suppose that $F$ has obtained 1 in his measurement on $S$. How is it possible that
$W$ has still a probability $|\alpha|^2$ of obtaining 0 on the same system, after the measurement of $F$ ?
This  ``paradox" is resolved operationally, since Wigner's measurement will never reveal 
any disagreement with his friend's measurement. 

In fact this situation illustrates the ``history collapse" effect of measurements, discussed in \cite{LC1,LC2}.
Consider the circuit version of Wigner's friend experiment in its simplest version, where
$S$ is a single qubit system and $F$ is a single qubit ``observer".  The evolutions (\ref{m0}), (\ref{m1}) can be realized just with a CNOT gate\footnote{The CNOT is a two-qubit gate acting on the computational basis as as $|0\rb|0\rb \rightarrow
|0\rb|0\rb$, $|0\rb|1\rb \rightarrow
|0\rb|1\rb$, $|1\rb|0\rb \rightarrow
|1\rb|1\rb$, $|1\rb|1\rb \rightarrow
|1\rb|0\rb$, the first qubit being the {\sl control} and the second being the {\sl target}.} where $S$ is the control and $F$ is the target qubit, and choosing $|init\rb = |0\rb$: 
\sk
\includegraphics[scale=0.45]{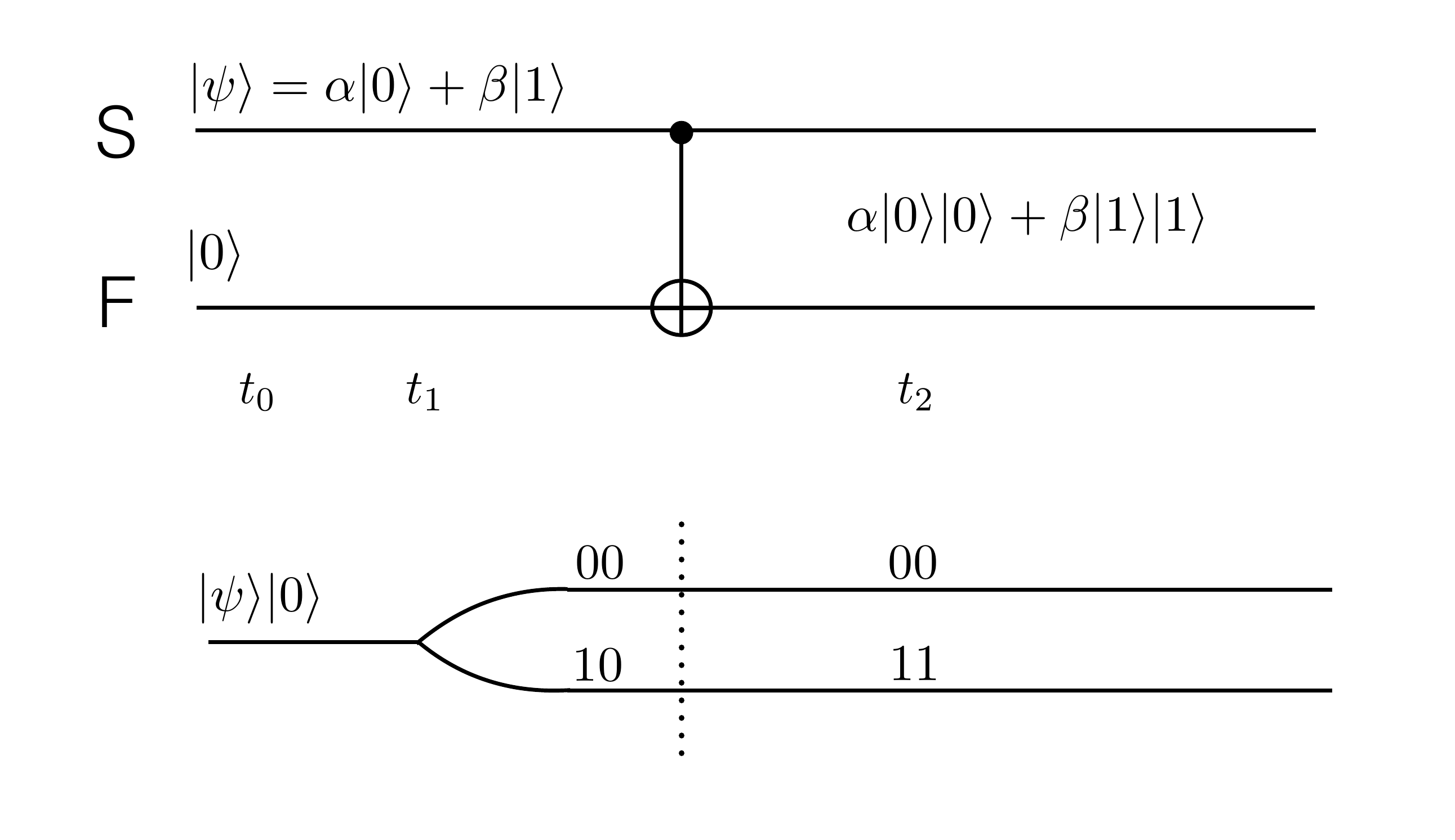}

\noi {\bf Fig. 1}  {\small Circuit describing the measurement of $F$ on $S$, as seen by $W$, and the corresponding history diagram. }  
\sk

\noi As discussed in \cite{LC2}, a quantum system can be described by
a {\sl history vector} $|\Psi\rb$ that encodes its whole time evolution:
\eq
|\Psi \rb = \sum_\gamma A(\gamma) |\gamma\rb \label{historyvector}
\en
where $\gamma$ are histories with nonvanishing amplitudes. A short summary follows, 
where we recall the definition of histories and corresponding amplitudes.
\sk
\noi By {\sl history} we mean a sequence of possible measurement results, at discrete times $t_1,t_2,...$ (time discretization is assumed for simplicity). The measured observables can be different at each $t_i$. With usual Born rules we can
compute the probability $p(\psi,\gamma)$ of obtaining a particular sequence $\gamma=\gamma_1,\gamma_2,...$ of results at times $t_1,t_2,...$, if the system starts in state $|\psi\rb$. We find
\eq
p(\psi,\gamma) = Tr (C_{\psi,\gamma} C_{\psi,\gamma}^\dagger ) \label{successive}
\en
where the chain operator $C_{\psi,\gamma}$ is defined as 
\eq
C_{\psi,\gamma} = P_{\gamma_{n}}  U(t_n,t_{n-1}) ~ P_{\gamma_{n-1}} ~ U(t_{n-1},t_{n-2})  \cdots P_{\gamma_{1}}~
U(t_1,t_0) P_{\psi} \label{chain}
\en
with $P_{\psi}=|\psi\rb \lb \psi|$; $P_{\gamma_{i}} $ are projectors on eigensubspaces of 
observables corresponding to the eigenvalues $\gamma_i$ and $U(t_{i+1},t_i)$ is the unitary evolution operator between times $t_i$ and $t_{i+1}$. If $\gamma_n$ is nondegenerate, the chain operator can be written as:
\eq
C_{\psi,\gamma} = |\gamma_n\rb A(\psi,\gamma) \lb \psi|
\en
the {\sl amplitude} $A(\psi,\gamma)$ being defined by
\eq
A(\psi,\gamma) = \lb \gamma_n| P_{\gamma_{n-1}} ~ U(t_{n-1},t_{n-2})  \cdots P_{\gamma_{1}}~
U(t_1,t_0) |\psi\rb  \label{amplitude}
\en
and the probability (\ref{successive}) is just the square modulus of the amplitude (\ref{amplitude}).
The generalization to a degenerate eigenvalue $\gamma_n$ is straightforward (see \cite{LC2}).
The vectors $|\gamma\rb$ in (\ref{historyvector}) are in 1-1 correspondence with all possible 
histories $\gamma_1,...\gamma_n$, and assumed to form an orthonormal basis. Probabilities
for (any sequences of) measurements can be easily computed using appropriate projections of the history vector (\ref{historyvector}) \cite{LC2}.
\sk
We apply now this formalism to the circuit that models Wigner's friend experiment, given in Fig.1. Note that 
it represents the system $S+F$ accessible only to $W$: the measurement of $F$ on the subsystem $S$ is represented by the unitary gate $CNOT$, and not by a projection.  Starting from an initial state $|\psi\rb|0\rb$ the history vector contains only two histories, since the only nonvanishing amplitudes 
are
\eqa
& & A(00,00) =\lb 00| CNOT |00\rb \lb 00| \psi,0\rb = \alpha \\
& & A(10,11) = \lb 11| CNOT |10\rb \lb 10| \psi,0\rb = \beta
\ena
Suppose now that  $W$ at time $t_2$ measures $F$ and obtains 1. This result is compatible only with the history $(10,11)$. Its probability is given the square modulus of the corresponding amplitude, i.e. $|\beta|^2$.
The history vector collapses to
\eq
|\Psi\rb = |10,11\rb
\en
Even if the measurement is performed at time $t_2$, the history involves also
values at time $t_1 < t_2$: in this sense the collapse ``modifies the past". Even if at $t_1$
$F$ had obtained 0 in his measurement of $S$, the measurement of $W$ ``undoes" this event
and realigns it to agree with his outcome at $t=t_2$. This ``undoing" of past events has 
been discussed by various authors (see for ex. \cite{brukner2}), and is really a matter of {\sl interpretation}, since there is no possibility of ``checking" the undoing. However note that history collapse is 
{\sl not} a matter of interpretation, if we require simultaneity of collapse in entangled
spacelike separated systems to be valid in every reference frame \cite{LC1}.
\sk
In conclusion, at first sight it may appear that Wigner ($W$) and his friend ($F$) give different descriptions of $F$'s measurement of $S$. In particular $F$'s description involves a collapse, whereas $W$ describes a unitary evolution. $W$ could detect interference effects in measurements on $S+F$, whereas $F$ sees no such effects after his measurement on $S$. But this difference is only due to $W$ and $F$ considering {\sl different} physical systems: $W$ describes $S+F$, whereas $F$ can only describe $S$. If we compare instead their descriptions {\sl of the same system} $S$, the difference  essentially vanishes as discussed above. The only residual difference is due to lack of information, {\sl not to
quantum mechanical effects}, and in our opinion does not motivate any ``relational" or subjective interpretation of quantum states.
\sk
{\bf Note: } in recent times Gedanken experiments \cite{FR,brukner1,bong} extending Wigner's friend setup have been proposed to probe the issue of observer independence. A real experiment has been carried out
\cite{proietti} in 2019, showing violation of Bell-type inequalities derived by assuming free-choice, locality and observer independence. However extreme care is necessary to uncover all assumptions made in real or Gedanken experiments. For example the analysis in ref. \cite{FR} has been criticized by
a number of authors \cite{araujo,sudbery,kastner}, and inconsistences have been pointed out.
It is therefore not clear to us that these experiments could directly support observer independence of quantum states or events.

\section*{Acknowledgements}

We thank Carlo Rovelli for correspondence on 
recent experiments.
This work is supported by the research funds of the Eastern Piedmont University and
INFN - Torino Section.

\vfill\eject
\end{document}